\documentclass[longauth, structabstract]{aa}  
\usepackage{natbib}
\usepackage{graphicx}
\usepackage{txfonts}
\usepackage{color}
\begin{document}
  \title{Search for an extended VHE gamma-ray emission \\from Mrk 421 and Mrk 501 with the MAGIC Telescope}

% authors 15.2.2010  Format A&A
%
\author{
 J.~Aleksi\'c\inst{1} \and
 L.~A.~Antonelli\inst{2} \and
 P.~Antoranz\inst{3} \and
 M.~Backes\inst{4} \and
 C.~Baixeras\inst{5} \and
 J.~A.~Barrio\inst{6} \and
 D.~Bastieri\inst{7} \and
 J.~Becerra Gonz\'alez\inst{8} \and
 W.~Bednarek\inst{9} \and
 A.~Berdyugin\inst{10}
% ...
}
\institute { IFAE, Edifici Cn., Campus UAB, E-08193 Bellaterra, Spain
 \and INAF National Institute for Astrophysics, I-00136 Rome, Italy
 \and Universit\`a  di Siena, and INFN Pisa, I-53100 Siena, Italy
 \and Technische Universit\"at Dortmund, D-44221 Dortmund, Germany
 \and Universitat Aut\`onoma de Barcelona, E-08193 Bellaterra, Spain
 \and Universidad Complutense, E-28040 Madrid, Spain
 \and Universit\`a di Padova and INFN, I-35131 Padova, Italy
 \and Inst. de Astrof\'{\i}sica de Canarias, E-38200 La Laguna, Tenerife, Spain
 \and University of \L\'od\'z, PL-90236 Lodz, Poland
 \and Tuorla Observatory, University of Turku, FI-21500 Piikki\"o, Finland
% ...
}

\offprints{corresponding authors J.~Sitarek (jsitarek@mppmu.mpg.de), R.~Mirzoyan (razmik@mppmu.mpg.de)}
  
%  \date{Received September 15, 1996; accepted March 16, 1997}
  \date{}
  
  \abstract
  {
    Part of the very high energy $\gamma$-ray radiation coming from extragalactic sources is absorbed through the pair production process on the extragalactic background light photons. 
Extragalactic magnetic fields
alter the trajectories of these cascade pairs and, in turn, convert
cosmic background photons to gamma-ray energies by inverse Compton
scattering.  
These secondary photons can form an extended halo around bright VHE sources.
}
  {We searched for an extended emission around the bright blazars Mrk 421 and Mrk 501  using the MAGIC telescope data.}
  {If extended emission is present, the angular distribution of reconstructed gamma-ray arrival directions around the source is broader than for a point-like source. 
In the analysis of a few tens of hours of observational data taken from Mrk~421 and Mrk~501 we used a 
newly developed method that provides better angular resolution. This method is based on the
usage of multidimensional decision trees.
Comparing the measured shapes of angular distributions with those expected from a point-like source one can 
detect or constrain possible extended emission around the source.
We also studied the influence of different types of systematic errors on the shape of the distribution of reconstructed gamma-ray arrival directions for a point source.}
  {We present upper limits for an extended emission calculated for both sources for various source extensions and emission profiles. 
We obtain upper limits on the extended emission around the Mrk~421 (Mrk~501) on the level of $<$ 5\% ($<$ 4\% ) of the Crab Nebula flux above the energy threshold of 300 GeV.
Using these results we discuss possible constraints on the extragalactic magnetic fields strength around a few times $10^{-15}$~G.
}
{}
%  {conclusions}
%{}

   \maketitle

\section{Introduction}

Blazars are well-known extragalactic sources of Very High Energy (VHE) $\gamma$-rays.
Their radiation traverses over large distances through the extragalactic space filled with CMB (Cosmic Microwave Background) and EBL (Extragalactic Background Light) photons and could be absorbed via the pair production process.
This effect can lead to a change in the spectral shape of observed radiation and it can be used to constrain the EBL density (see e.g. \cite{stecker1992}, \cite{eblpaper}). 

A possible extended emission around extragalactic sources of VHE $\gamma$-rays  was first discussed by \citet{aharonian_halo}.
$\gamma$-rays with energies larger than 10 TeV are strongly absorbed via a pair production process on EBL and CMB photons relatively close to the source.
Secondary $\gamma$-rays can be produced in a cascade initiated by those primary photons.
Note that the redshift dependent energy for which the gamma-ray opacity is equal 1 vary within a factor of $\sim 3$ in different EBL models \citep{Kneiske:2003tx, Stecker:2005qs, Franceschini:2008tp, gilmore08, Primack:2008nw, finke10}. 

Magnetic fields are very non-uniform and their strength can vary within many orders of magnitude depending on the location in the large scale structure.
So far they have been measured only in the galaxies with strength $\sim\mu$G \citep{kz08, beck08}, the cores of galaxy clusters (within the inner 100 kpc) $\sim$ several $\mu$G \citep{ct02}, and near borders of few clusters with strength $10^{-8}$ -- $10^{-7}$~G (on Mpc scales) \citep{xkhd06,kksp07}. 
On larger distance scales the magnetic field strengths are not known but they have to be much weaker (see e.g. \cite{angelis}).
Moreover, the theoretical models predict very weak ($B  \ll 10^{-12}$~G) extragalactic magnetic fields (EGMF) in voids in large scale structures, outside galaxies and galaxy clusters \citep{kronberg94,grasso00,widrow2002,neronov09}. 
It is believed that voids make up a significant part of the space volume. 

Development of a secondary gamma-ray cascade depends on its original energy and the strength of surrounding magnetic field. 
Two cases can be considered. 
First, if the TeV source is located in an intense magnetic field region and the maximal gamma-ray energy is large ($E_{max}\gg 50$ TeV), gamma-rays would produce $e^+e^-$ pairs directly near the source (within a few Mpc). 
These will be isotropized in the strong magnetic field which might exist around the source host galaxy or galaxy cluster \citep{aharonian_halo}.

Second, the mean free path for gamma-rays of moderate energies ($E<50$~TeV) is longer, therefore they can traverse larger distances from the source. 
Therefore $e^+e^-$ pairs will be produced in regions with much weaker magnetic fields \citep{plaga95, neronov07, neronov2, kachelriess}.
In this case the deflections of the pair trajectories by EGMF are not large enough to make the secondary cascade emission isotropic.
Instead the inverse Compton scattering of CMB photons by $e^+e^-$ pairs will produce secondary $\gamma$-rays with a slightly different direction with respect to the primary photons.
The re-direction of the cascade photons into the field of view of the telescope can lead to the appearance of an extended emission around the point-like source, even in the case $B\le 10^{-12}$~G \citep{neronov07,neronov2,kachelriess}.
In fact the extended emission would be produced between the source and the observer.
For a distant observer that extended emission will appear as superimposed onto the point source thus mimicking a halo.

The secondary $\gamma$-rays also can initiate cascades, provided that the optical depths in the EBL radiation field are still large enough.
The energies of further generation of $\gamma$-rays may be below the energy threshold of VHE $\gamma$-ray instruments.

Blazars are known to be strongly variable in particular at the $\gamma$-ray energies (e.g. \cite{aharonian_pks}, \cite{quantum_gravity}).
The extended emission component cannot follow the original time profile of the emission.
This is because the secondary cascade photons do not propagate along the same path as the direct $\gamma$-rays from the primary source. 
Instead it will be delayed and stretched to much longer time scales (up to $\sim10^6$ years at $\sim 100$~GeV energy for $B\sim 10^{-12}$~G, \cite{neronov07,neronov09}, see also \citet{plaga95,dai02,murase08}).
Thus the direct emission from the point source will be overlayed on the extended emission component. 
The latter can constitute a part of the ``quasi'' constant, quiescent emission. 

The first attempt to detect extended emission around extragalactic TeV $\gamma$-ray sources was performed by the HEGRA instrument for Mrk~501. It yielded only an upper limit of 5-10\% of the Crab Nebula flux (at energy $\geq 1$ TeV) on angular scales of 0.5 to 1$^\circ$, \citep{hegra_halo}.

In this paper we report on our search for extended emission of VHE $\gamma$-rays from the bright blazars Mrk~421 and Mrk~501 using the 17m diameter MAGIC Imaging Atmospheric Cherenkov Telescope (IACT). 
The details of the telescope, its performance and the standard analysis chain
are described in \cite{crab_paper}.
Due to the large size of the mirror dish and improved light sensors, the MAGIC trigger threshold ($\sim 50$~GeV) is $\sim 2-3$ times lower compared to other operating IACTs.

This paper is structured in the following way: 
In section 2. we describe the analysis method, which was used for searching for an extended emission. 
Also, we present a new method that improves the angular resolution.
In  section 3. we perform analyses of possible systematic effects which can mimic the existence of an extended emission. 
Then, in section 4. we describe the data sample used for the analysis. 
In section 5. we present results of the analysis: the upper limits on the extended emission from Mrk~421 and Mrk~501. 
Finally, in section 6. we discuss how those upper limits constrain the strength of the EGMF.
This is followed by a short discussion on future prospects and conclusions.

\section{Analysis method}
We parameterize the images of air showers by using the so-called Hillas parameters \citep{hillas}. 
The angular distance between the center of gravity of the image and the shower direction (so called DISP) is correlated with geometrical and timing properties of the image. 
The arrival direction for every event can be estimated. 
The distribution of the squared angular distance between the estimated and the true source position ($\theta^2$) is narrow and has a peak at $\theta=0$ for a point-like $\gamma$-ray source. 
In case of an extended source this distribution shall be broader.

The DISP parameter is proportional to the ellipticity ($1-Width/Length$) of the image. 
By including the dependence on the possible truncation of the image at the edge of the camera and also on the parameter $Size$ (sum of total charge of an image), one can improve the precision of DISP (see \cite{disp_paper}). 

\subsection{The novel Random Forest DISP method}
For this study we developed a novel method for the DISP estimation that along with standard Hillas parameters ($Width$, $Length$, $Size$) includes the $Leakage$, the $Time\ Gradient$, and the dependence on the zenith angle of observations.
$Leakage$ is a measure of a truncation of images due to the camera edge effects.
It is defined as the ratio of the charge in the last two rings of pixels in the camera to the total charge.
The parameter $Time\ Gradient$ is defined as the derivative of signal arrival time in pixels along the main axis of the image.
For a given event the $Time\ Gradient$ is strongly correlated with the impact parameter of the parent shower. 
Since for a given zenith angle and $\gamma$-ray energy the DISP is a simple function of the impact parameter, one could expect to improve the DISP estimation, and as a result also the angular resolution by using the fast timing properties of the image. 
The MAGIC telescope's ultra-fast time response because of the parabolic reflector shape, special PMTs and 2 GSample/s FADC readout enabled us to include the $Time\ Gradient$ parameter in our analysis thus enhancing the sensitivity of the telescope \citep{timepaper}.

\begin{figure}%%%%%%%%%%%%%%%%%%%%%%%%%%%%%%%%%%
  \centering
\includegraphics[width=8cm]{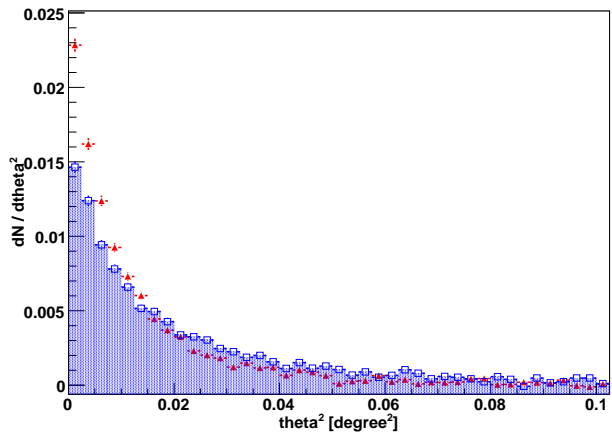}
\caption{Comparison between the $\theta^2$ distributions of the $\gamma$-ray excess obtained from 43h of Crab Nebula data with the parametrized DISP value (blue shaded area), and the RF DISP method (red triangles).
}\label{fig_rf_vs_par}
\end{figure}

In order to combine information from both the geometrical and timing properties of images in the most efficient way, we used multidimensional decision trees - the so-called Random Forest (RF) method. 
It is widely used for the $\gamma$/hadron separation and the energy estimation \citep{bock2004, RF}. 
A comparison between the novel RF method (hereafter RF DISP) and the standard parametrized DISP is presented in fig.~\ref{fig_rf_vs_par}.
% bw
%% \begin{figure}%%%%%%%%%%%%%%%%%%%%%%%%%%%%%%%%%%
%%   \centering
%% \includegraphics[width=8cm]{rf_vs_par.eps}
%% \caption{Comparison between the $\theta^2$ distributions of the $\gamma$-ray excess obtained from 43h of Crab Nebula data with the parametrized DISP value (empty squares), and the RF DISP method (triangles).
%% }\label{fig_rf_vs_par}
%% \end{figure}%%%%%%%%%%%%%%%%%%%%%%%%%%%%%%%%%%
% color
%%%%%%%%%%%%%%%%%%%%%%%%%%%%%%%%%%
The RF DISP provides a substantially narrower $\theta^2$ distribution and improves the angular resolution (defined as the 40\% containment radius for a point like source, equivalent to one standard deviation of a two dimensional gaussian distribution) by $\sim 20-30\%$.
This improvement is due to the usage of a) Random forest method instead of simple parameterization and b) $Time\ Gradient$ information. 
This enhances the telescope performance for the search for an extended emission.

%The shape and the width of a $\theta^2$ distribution for a point-like source depends on many factors, the important being the energy of the shower. 
The shape and the width of a $\theta^2$ distribution for a point-like source depends on many factors, among them the energy of the showers. 
For a higher energy shower, due to the large number of particles in the shower maximum, one has a higher signal-to-noise ratio, and the resulting image has more precisely defined parameters. %and correspondingly  image has better defined parameters.
For this reason we have selected events with large $Size$ (the total measured charge of the image), which improves the precision of the reconstruction of the shower direction. 
%In our analysis we used only showers with $Size>400$ photoelectrons, that provide a relatively high precision of the $\theta^2$ determination.
In our analysis we used only showers with $Size>400$ photoelectrons which allows us to determine the $\theta^2$ with relatively high precision.
This leads to an energy threshold (defined as the peak of the Monte Carlo simulated differential energy distribution) of 300 GeV.

%% black and white figure
%% \begin{figure}%%%%%%%%%%%%%%%%%%%%%%%%%%%%%%%%%%
%%   \centering
%%  \includegraphics[width=8cm]{mc_ext_comp.eps}
%%  \includegraphics[width=8cm]{mc_ext_diff.eps}
%%   \caption{
%% Comparison of Monte Carlo cumulative $\theta^2$ distributions (upper panel) for a source with 80\% point-like and 20\% extended emission and difference between them (lower panel). 
%% The characteristic radius of the extension is equal to $0.1^\circ$ (dotted), $0.2^\circ$ (dashed) or $0.3^\circ$ (dot-dashed).
%% The extended part of the emission is simulated as a flat disc ($dN/d\theta^2=\mathrm{const}$).
%% The purely point-like source is shown as solid line.
%% A random mispointing up to 0.03$^\circ$ has been included in the simulations.
%% }
%%   \label{fig_th2cum}
%% \end{figure}%%%%%%%%%%%%%%%%%%%%%%%%%%%%%%%%%%
%% color figure
\begin{figure}%%%%%%%%%%%%%%%%%%%%%%%%%%%%%%%%%%
  \centering
 \includegraphics[width=8cm]{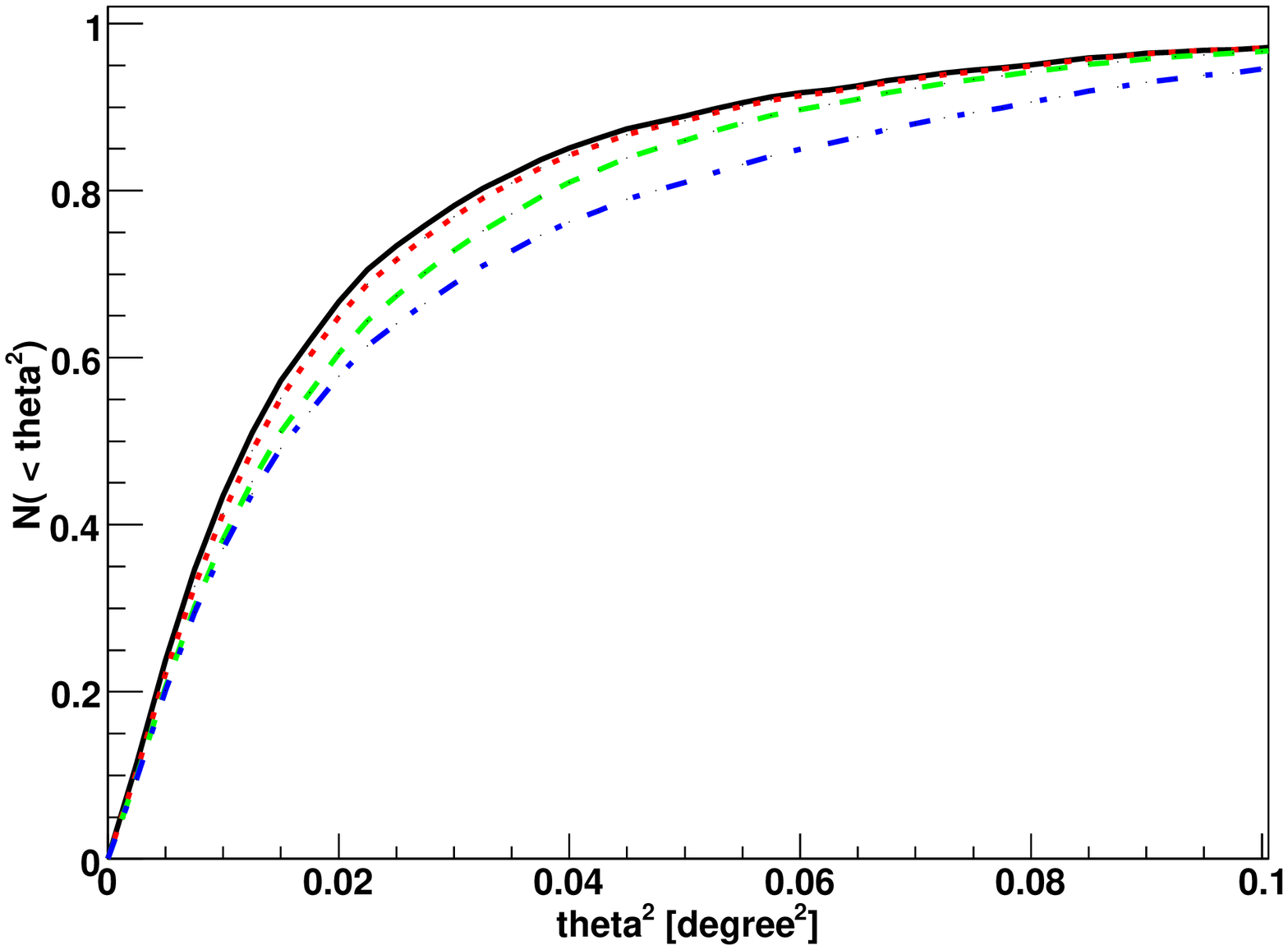}
 \includegraphics[width=8cm]{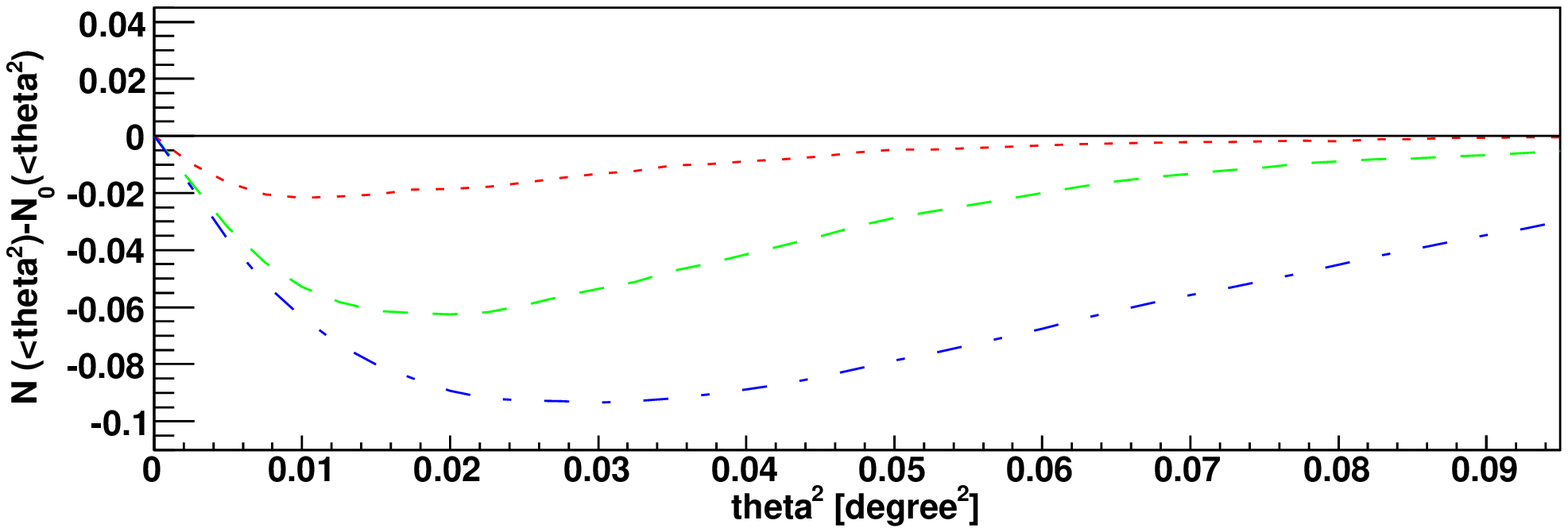}
  \caption{
Comparison of Monte Carlo cumulative $\theta^2$ distributions (upper panel) for a source with 80\% point-like and 20\% extended emission and difference between them (lower panel). 
The characteristic radius of the extension is equal to $0.1^\circ$ (red), $0.2^\circ$ (green) or $0.3^\circ$ (blue).
The extended part of the emission is simulated as a flat disc ($dN/d\theta^2=\mathrm{const}$).
The purely point-like source is shown as black line.
A random mispointing up to 0.03$^\circ$ has been included in the simulations.
}
  \label{fig_th2cum}
\end{figure}%%%%%%%%%%%%%%%%%%%%%%%%%%%%%%%%%%

Let us consider a situation where the excess observed from a hypothetical source is a mixture of a point source and a weak extended emission with a given profile.
The cumulative $\theta^2$ distributions for point-like and extended sources are shown in fig.~\ref{fig_th2cum}.
Using this figure, the angular resolution for this analysis is estimated to be $\lesssim 0.1^\circ$.

\subsection{Source extension}
If the characteristic extension $\theta_{ext}$ is much smaller than the telescope's PSF, the distributions of a purely point-like and a partially extended source are very similar. 
For larger extensions clear differences between the distribution shapes can be seen (see fig.~\ref{fig_th2cum}).

To investigate possible extended emission from blazars we adopted a method similar to the one used in \cite{hegra_halo}. We calculated the ratio of event rates in two $\theta$ ranges: 
%\begin{equation}
$f= \frac{N(\theta_1<\theta<\theta_2)}{N(0<\theta<\theta_1)},$
%\end{equation}
and compared them with the ratio calculated for a point-like source.
The $\theta_1$ and $\theta_2$ values are calculated with the help of Monte Carlo simulations searching for the most significant difference between a purely point source and a source with an extended emission for a given size and profile of extension. 
For every considered flux, radius and profile of the extended emission we calculated the value of $f$ and of the corresponding significance of the extension with MC simulations. 

To minimize the systematic errors in modeling a point source we have selected a data sample taken from the Crab Nebula (normalized to the Mrk~421 or Mrk~501 flux).
The extension of the Crab Nebula in the VHE gamma-rays, as shown in \citet{hegra_crab} is below $0.025^\circ$, which makes it essentially a point-like source for MAGIC.
Also the mean spectral slopes of both Mrk~501 and Mrk~421 in the considered data sample are rather similar to that of the Crab Nebula.

\section{Analysis of the systematic effects}
Systematic effects can degrade the precision of the estimation of the arrival direction of $\gamma$-rays.
Some of those effects, if not taken into account properly, can emulate an extended emission around the point-like source.
Here we present a study of those effects.

\subsection{Optical PSF}
The optical point spread function (PSF) of the telescope smears out images. 
This increases $Width$ and $Length$ of any given image, thus changing its ellipticity. 
Therefore the DISP method will be affected and the $\theta^2$ distribution will become broader.
Variation of the optical PSF will be reflected in the measured angular resolution. The optical PSF varies across the field of view of the parabolic reflector of MAGIC. 
Also, varying gravitational loads of the mirror dish during observations produce small deviations  of the optical PSF, which are corrected by active mirror control system of MAGIC. 
The largest observed variations of the PSF of MAGIC are in the range of 20 \%.
% bw
%% \begin{figure}%%%%%%%%%%%%%%%%%%%%%%%%%%%%%%%%%%
%%   \centering
%%   \includegraphics[width=8cm]{mc_psf_th2.eps}
%%   \includegraphics[width=8cm]{mc_psf_th2_diff.eps}
%%   \caption{Monte Carlo cumulative $\theta^2$ distributions for a point-like source for two different values of the optical PSF of the reflector: $\sigma=0.036^\circ$ (solid), $\sigma=0.044^\circ$ (dashed) (upper panel), and a difference between them (lower panel).}
%%   \label{fig_psfs}
%% \end{figure}%%%%%%%%%%%%%%%%%%%%%%%%%%%%%%%%%%
%
% color
\begin{figure}%%%%%%%%%%%%%%%%%%%%%%%%%%%%%%%%%%
  \centering
  \includegraphics[width=8cm]{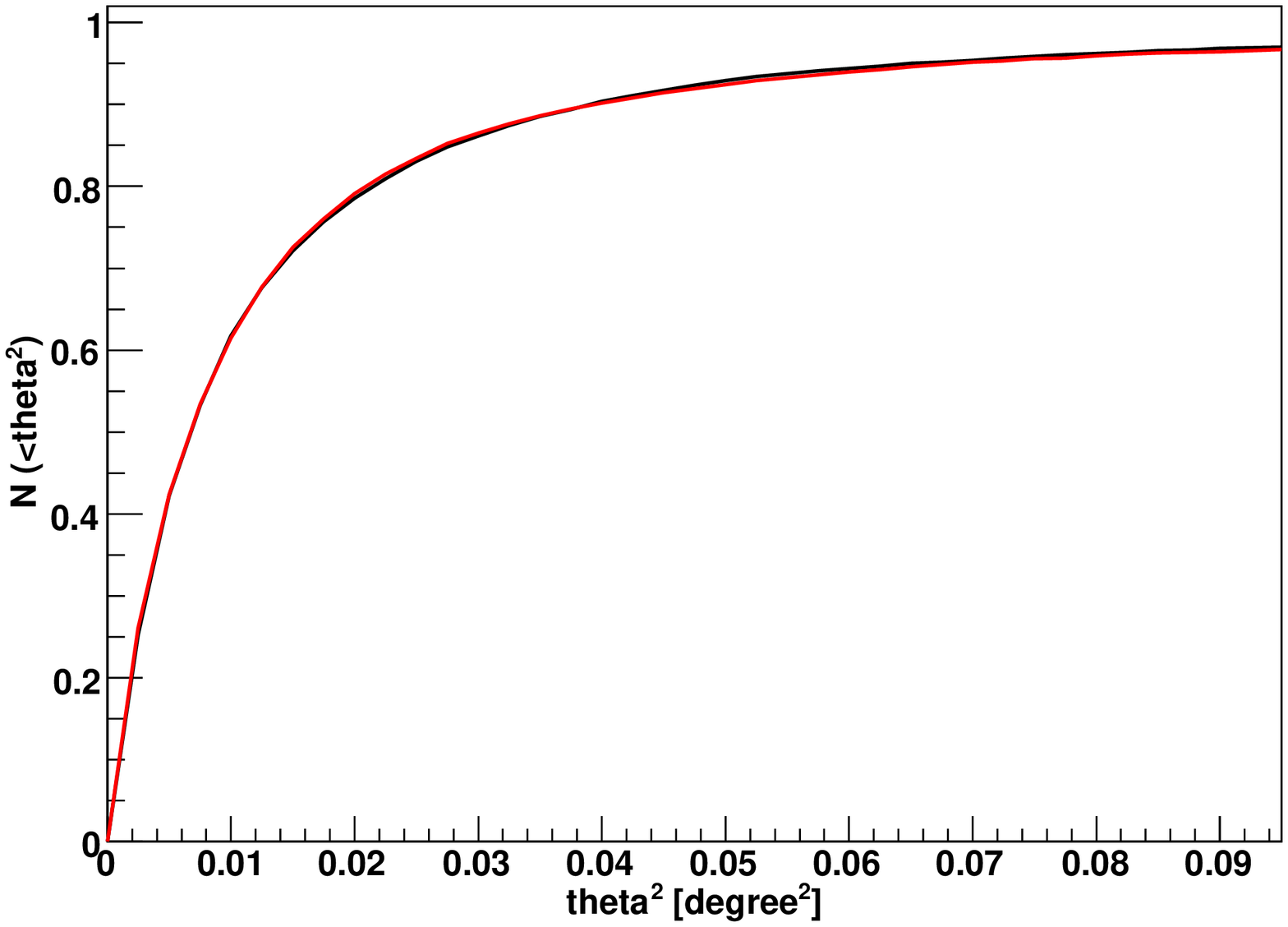}
  \includegraphics[width=8cm]{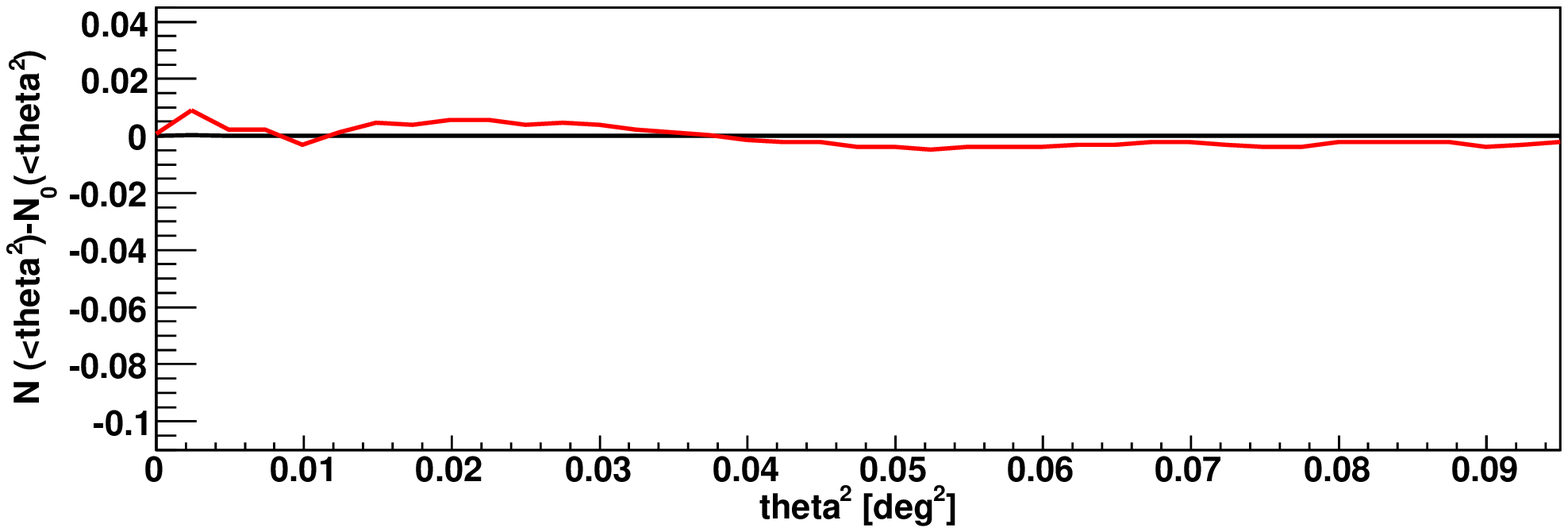}
  \caption{Monte Carlo cumulative $\theta^2$ distributions for a point-like source for two different values of the optical PSF of the reflector: $\sigma=0.036^\circ$ (black), $\sigma=0.044^\circ$ (red) (upper panel), and a difference between them (lower panel).}
  \label{fig_psfs}
\end{figure}%%%%%%%%%%%%%%%%%%%%%%%%%%%%%%%%%%
In fig.~\ref{fig_psfs} we show the result of a study of the influence of optical PSF on the shape of the $\theta^2$ distribution.
An increase of the optical PSF by e.g. 20\% is equivalent to $\lesssim$ 2\% admixture of an extended source with a characteristic extension radius of 0.2$^\circ$. 
We conclude that this effect is negligible for our study. 

\subsection{Spectral index}
As mentioned before, the $\theta^2$ estimation is more precise for higher energy $\gamma$-rays. 
This means that a source with a harder spectrum will have a more peaked $\theta^2$ distribution than a source with a softer spectral index.

% bw
%% \begin{figure}%%%%%%%%%%%%%%%%%%%%%%%%%%%%%%%%%%
%%   \centering
%% \includegraphics[width=8cm]{mc_slopes_th2.eps}
%% \includegraphics[width=8cm]{mc_slopes_th2_diff.eps}
%%   \caption{Monte Carlo cumulative $\theta^2$ distributions for a point-like source with a different spectral index: $-2.2$ (dashed), $-2.4$ (solid) $-2.6$ (dotted) (upper panel), and a difference between them (lower panel).}
%%   \label{fig_slopes}
%% \end{figure}%%%%%%%%%%%%%%%%%%%%%%%%%%%%%%%%%%
% color
\begin{figure}%%%%%%%%%%%%%%%%%%%%%%%%%%%%%%%%%%
  \centering
\includegraphics[width=8cm]{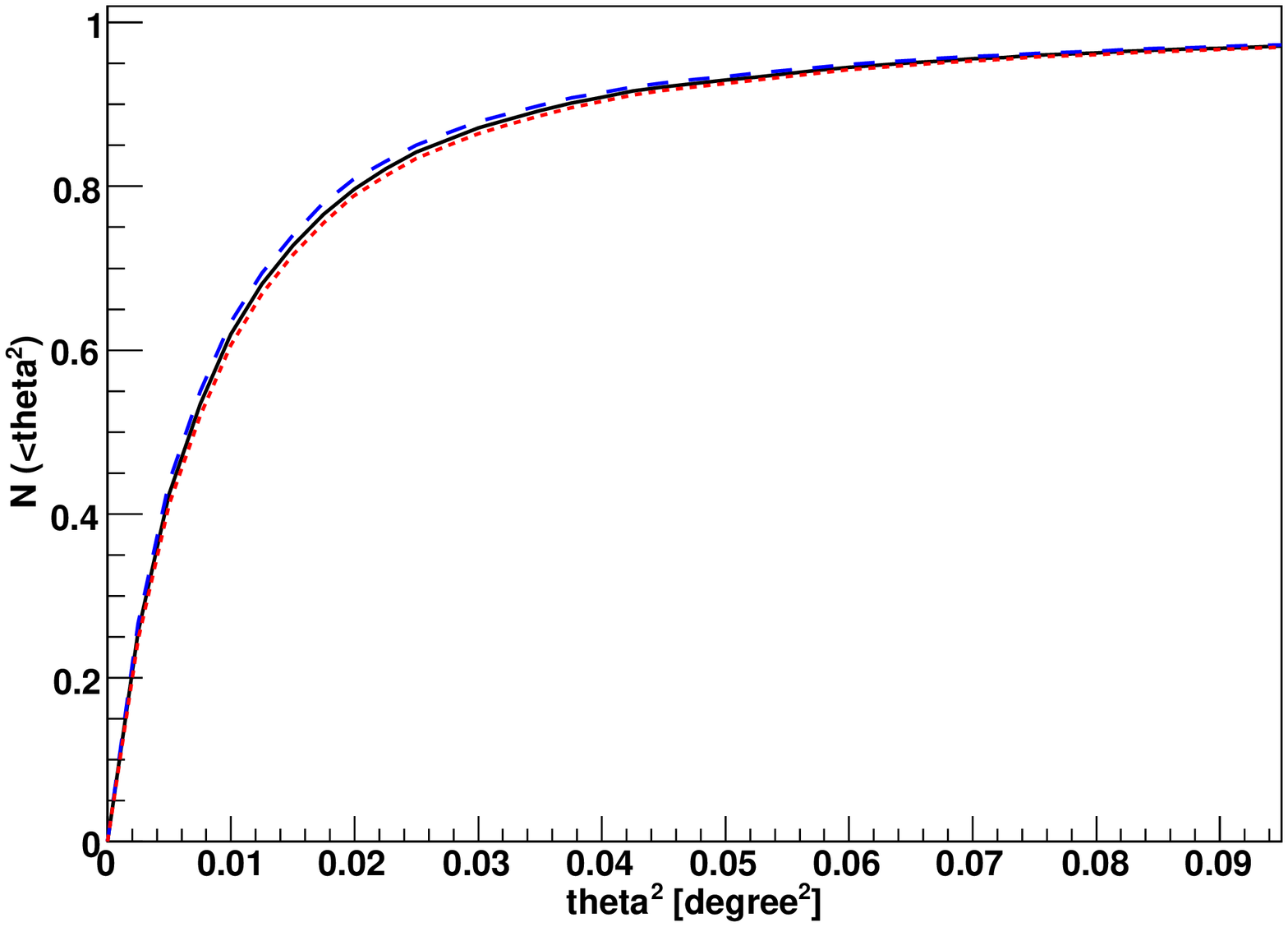}
\includegraphics[width=8cm]{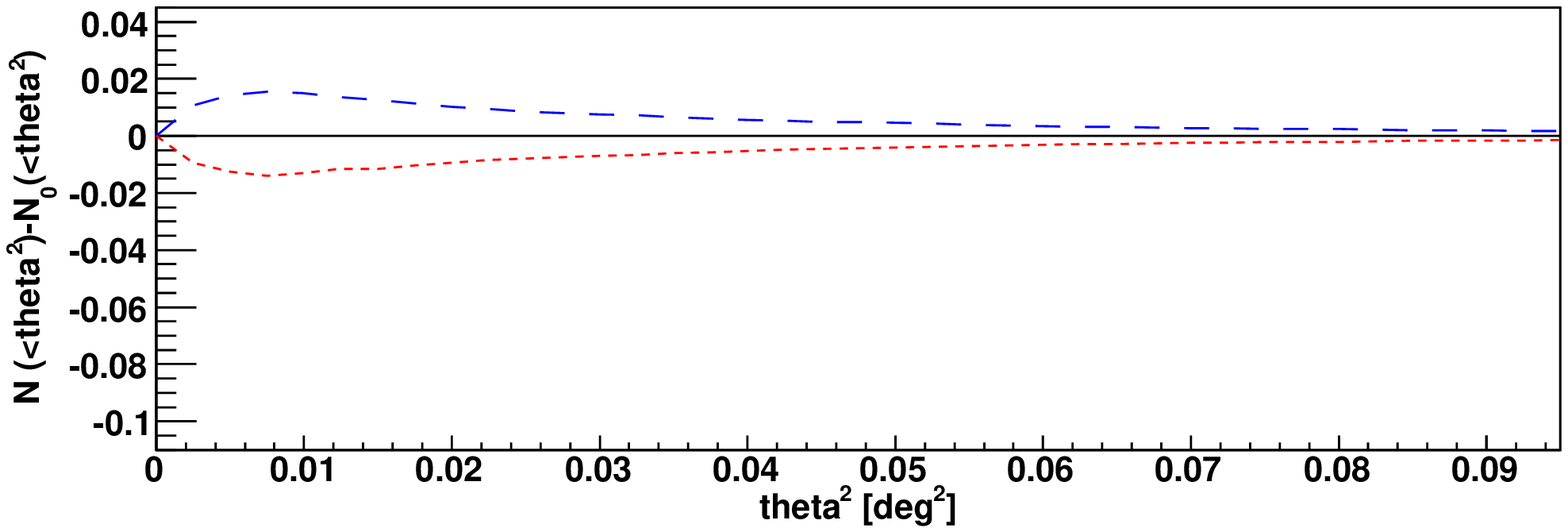}
  \caption{Monte Carlo cumulative $\theta^2$ distributions for a point-like source with a different spectral index: $-2.2$ (blue), $-2.4$ (black) $-2.6$ (red) (upper panel), and a difference between them (lower panel).}
  \label{fig_slopes}
\end{figure}%%%%%%%%%%%%%%%%%%%%%%%%%%%%%%%%%%

In fig.~\ref{fig_slopes} we present a comparison of cumulative $\theta^2$ distributions for spectral indices $-2.2$, $-2.4$ and $-2.6$.
The broadening of the $\theta^2$ distribution that corresponds to a steepening of the spectral index by 0.2 (a typical systematic error in spectral index determination) is comparable to having a 3\% admixture of extended emission with a characteristic extension of 0.2$^\circ$.
Compared to the statistical errors of this study, this effect is small.

\subsection{Mispointing}
A factor which limit the ability of IACTs to distinguish between point-like and extended sources is the possible mispointing of the telescope. 
Using a strong source like the Crab Nebula or Mrk~421 one can estimate the source position and then calculate the mispointing as a difference between the true and the estimated positions.
% bw
%% \begin{figure}%%%%%%%%%%%%%%%%%%%%%%%%%%%%%%%%%%
%%   \centering
%%   \includegraphics[width=8cm]{mispoint.eps}
%%   \caption{Difference between the true and the estimated source positions for Mrk~421 (squares) and Crab Nebula (triangles). 
%% Circles with radii $0.02^{\circ}$ and $0.03^\circ$ show the characteristic mispointing scale}
%%   \label{fig_mispoint}
%% \end{figure}%%%%%%%%%%%%%%%%%%%%%%%%%%%%%%%%%%
%color
\begin{figure}%%%%%%%%%%%%%%%%%%%%%%%%%%%%%%%%%%
  \centering
  \includegraphics[width=8cm]{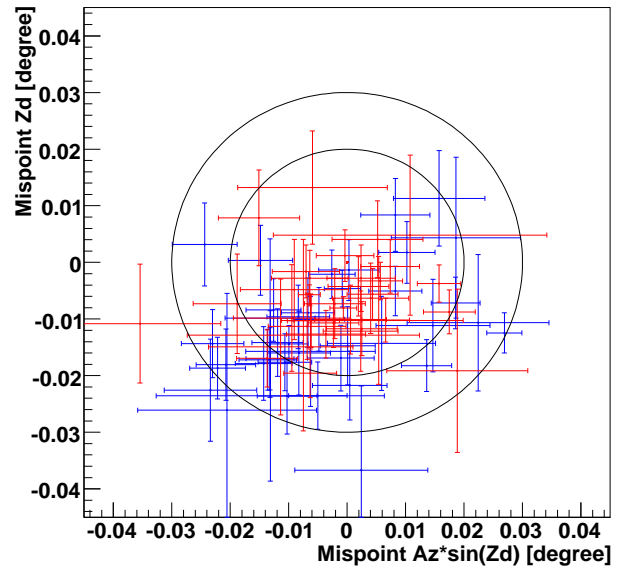}
  \caption{Difference between the true and the estimated source positions for Mrk~421 (red) and Crab Nebula (blue). 
Circles with radii $0.02^{\circ}$ and $0.03^\circ$ show the characteristic mispointing scale}
  \label{fig_mispoint}
\end{figure}%%%%%%%%%%%%%%%%%%%%%%%%%%%%%%%%%%
As it is shown in fig.~\ref{fig_mispoint} the accuracy of the source position reconstruction is  $\lesssim 0.02-0.03^\circ$ for these observations.
This is in agreement with our previous studies \citep{crab_paper}.
Those values are well below the angular resolution of the telescope and the investigated extensions of the AGN halos.

\section{Observations}
MAGIC, consisting of two 17m diameter telescopes, is located on the Canary island of La Palma at the Roque de los Muchachos Observatory (at 2200 m.a.s.l).
The data presented in this paper has been taken with the first MAGIC telescope.

Mrk 421 and Mrk 501 are nearby blazars and well-known VHE $\gamma$-ray sources.
Their spectra have been measured up to $\sim 10-20$ TeV (e.g. \cite{hegra_mrk501}, \cite{whipple_mrk421}).
At those energies absorption in EBL radiation fields becomes important so one may expect an extended emission component  due to an AGN halo.
Both sources are strongly variable. 
They are being monitored by various VHE $\gamma$-ray experiments for nearly 20 years. 
During this time both quiescent states with flux as low as  $\sim 0.15$ C.U. (Crab unit) and giant flares with flux up to $\sim 10$ C.U. were observed.

During 1995-1999 (with the exception of short time flares) the flux registered by the HEGRA instrument from Mrk 421 was well below 1 C.U. 
In 2000 it increased to a level $\sim 1$ C.U., and it further increased up to $\sim 2.5$ C.U. in 2001 (\cite{hegra_mrk421}).
Observed flux could be described using power-law with a spectral index $-2.4$ and an exponential cut-off at the energy of 3.4 TeV. 
In MAGIC observations of Mrk~421 performed between November 2004  and April 2005 the flux varied between $0.5-2$ C.U.. 
Also the cut-off energy in this time period seems to be at lower value $\sim 1.4$ TeV (\cite{magic_mrk421}).
In the low state the source seems to have a steeper spectrum with a spectral index $\sim -3$ between 0.5 and 7~TeV \citep{aharonian2002, hegra_mrk421} .

Observations of Mrk 501 are equally interesting. 
Historically the strongest activity period for this source was in 1997. 
The mean flux observed by the HEGRA, the Whipple and the CAT telescopes in that year was 1.3 - 3 C.U. 
The same instruments observed Mrk~501 in 1998-1999 in a low state of $\sim 0.15$ C.U. 
The source became more active in 2000 resulting in a flux measured by the HEGRA and Whipple instruments at a level of $0.35 - 1.2$ C.U. (see review of all those observations in \cite{magic_mrk501_2007}).
MAGIC observations of Mrk 501 in 2005 resulted in a mean flux of $\sim 0.5$ C.U. (\cite{magic_mrk501_2007}), 
while observations performed in 2006 show a low state at $\sim 0.2$ C.U. with a spectral index $-2.8$ (\cite{magic_mrk501_2009}).
The spectra of Mrk 501 can be well described by a (possibly curved) power-law (\cite{magic_mrk501_2007}).

For studing the possible extended VHE gamma-ray emission we selected recent MAGIC observational data from Mrk~421 and Mrk~501.  
To minimize systematic errors, only data taken at low ($<30^\circ$) zenith angles were used. 

Data from Mrk~501 have been taken in April/May 2008 in the so-called ON/OFF mode (where the source is in the center of the camera). 
After quality cuts 26h of ON data were selected. 
50h of OFF data were used for the background estimation. 
Mrk~501 was in a rather low state during the above-mentioned time period (mean flux in the sample $\sim 15\%$ C.U.).
The entire spectrum can be well fitted with a single power law with a spectral index of $-2.42\pm 0.03_{\rm stat}\pm 0.2_{\rm syst}$.
After correcting for the absorption due to the EBL, by using the \citet{Franceschini:2008tp} model, we obtain the source spectrum with an index of $-2.24\pm 0.03_{\rm stat}\pm 0.2_{\rm syst}$.

The Mrk~421 data was collected between December 2007 - February 2009. 
After quality cuts 38h of data were selected. 
The data was taken in the so called wobble mode, where the source position was shifted $0.4^\circ$ from the center of the camera. 
The opposite (with respect to the camera center) position was used for the background estimation. 
Since the background data was taken simultaneously with the source data, the systematic errors are small. 
A disadvantage of this approach is the fact that if the source extension is as large as $\sim 0.4^\circ$, the signal and the background regions start overlapping.
This effect, which can reduce the sensitivity, has been studied and included in the Monte Carlo simulations.
In the analyzed data sample Mrk~421 was in a high state ($\sim 1.3$ C.U.). 
Its spectrum in the wide energy range is best fitted with a flat (spectral index $-2$) power law with an exponential break point at the energy $E_{cut}=2.1$~TeV).
In the limitted energy range of interest (0.3 -- 3~TeV) the spectrum from this data sample can be also fitted with an effective power law with a spectral index of $-2.42\pm 0.02_{\rm stat}\pm 0.2_{\rm syst}$.

The $\theta^2$ distribution for a point like source was calculated by using the Crab Nebula data.
We use 43h of wobble mode Crab Nebula data, taken between October 2007 and March 2009, and 17h of ON mode data taken during December 2007 and January 2008.
This data-set can be described by a spectrum of $-2.3\pm 0.02_{\rm stat}\pm 0.2_{\rm syst}$ for the energy band (0.3 -- 3~TeV).

\section{Results}
\begin{figure}%%%%%%%%%%%%%%%%%%%%%%%%%%%%%%%%%%
  \centering
  \includegraphics[width=8cm]{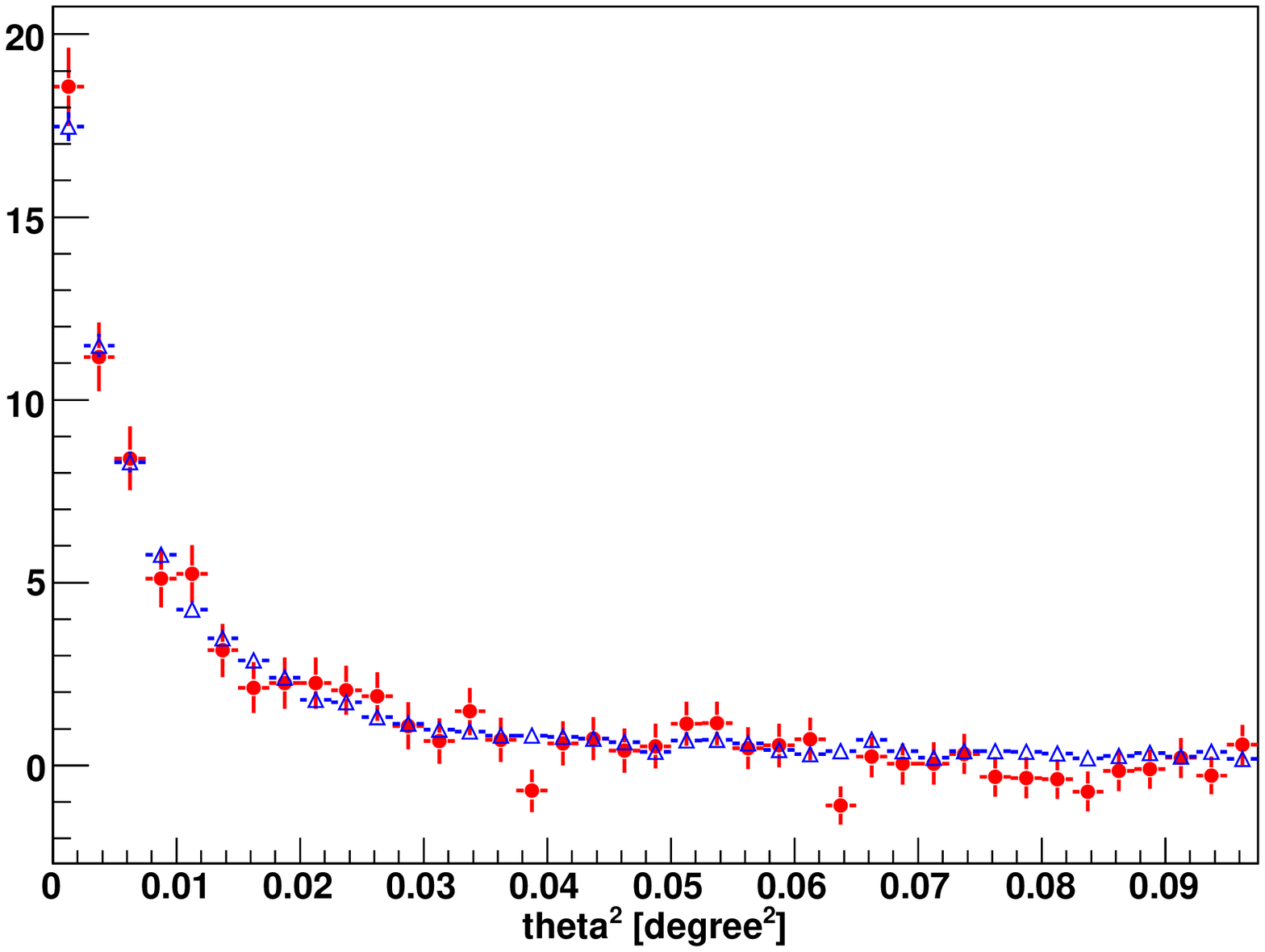}
  \includegraphics[width=8cm]{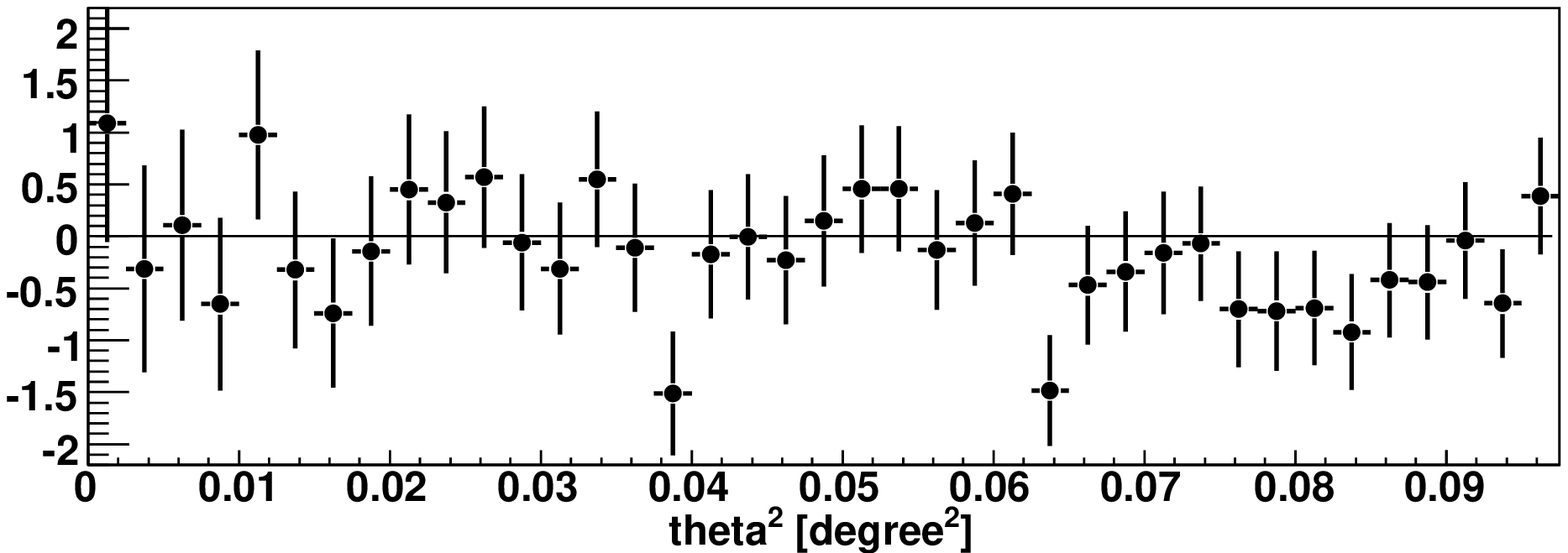}
  \caption{Comparison of the excess $\theta^2$ distribution for 26h of data from Mrk~501 (red circles) and a point-like source (blue triangles, 17h of Crab Nebula data) (upper panel) and the difference of both distributions (lower panel).}
  \label{fig_mrk501}
\end{figure}

\begin{figure}
  \centering
  \includegraphics[width=8cm]{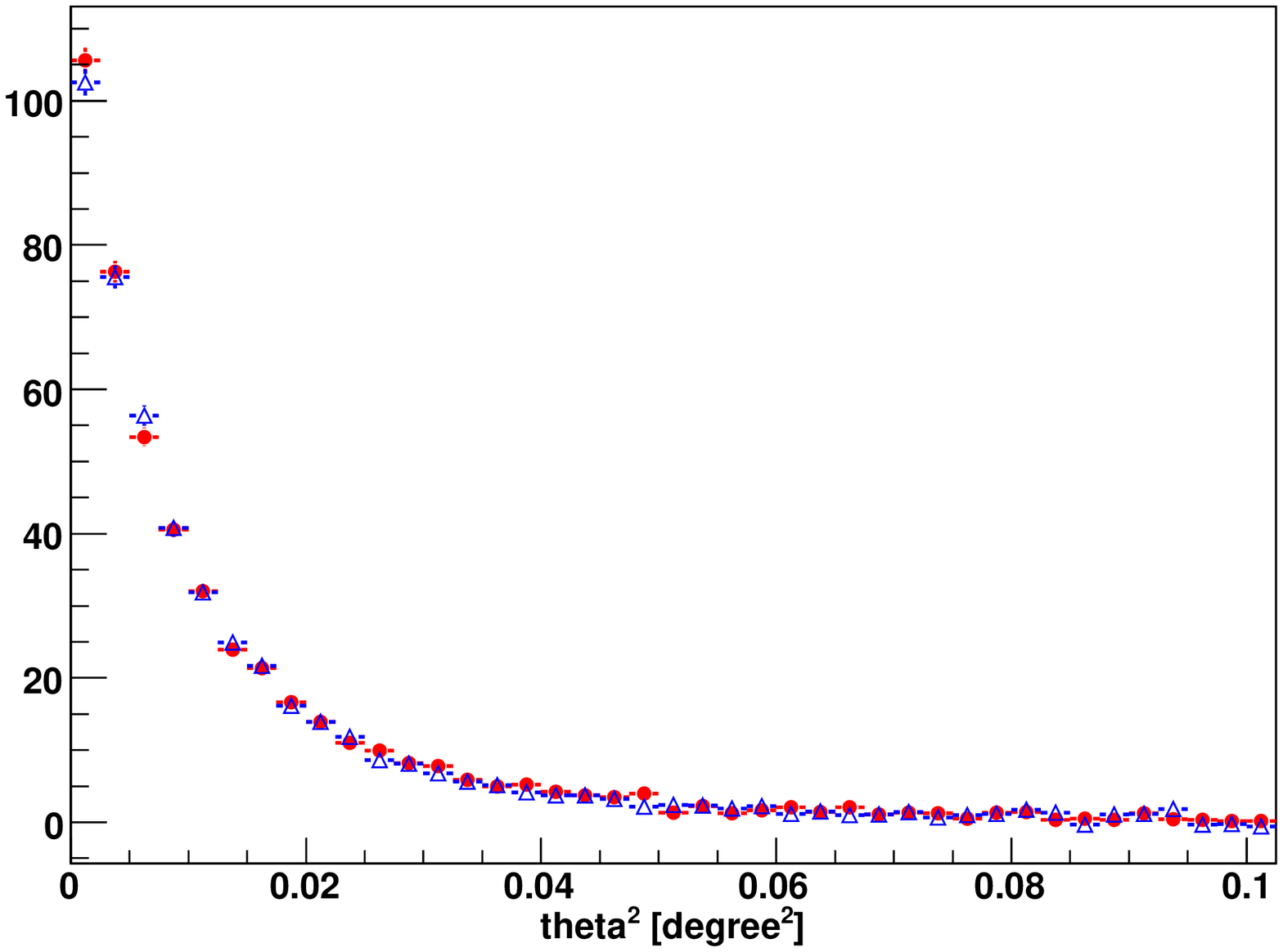}
  \includegraphics[width=8cm]{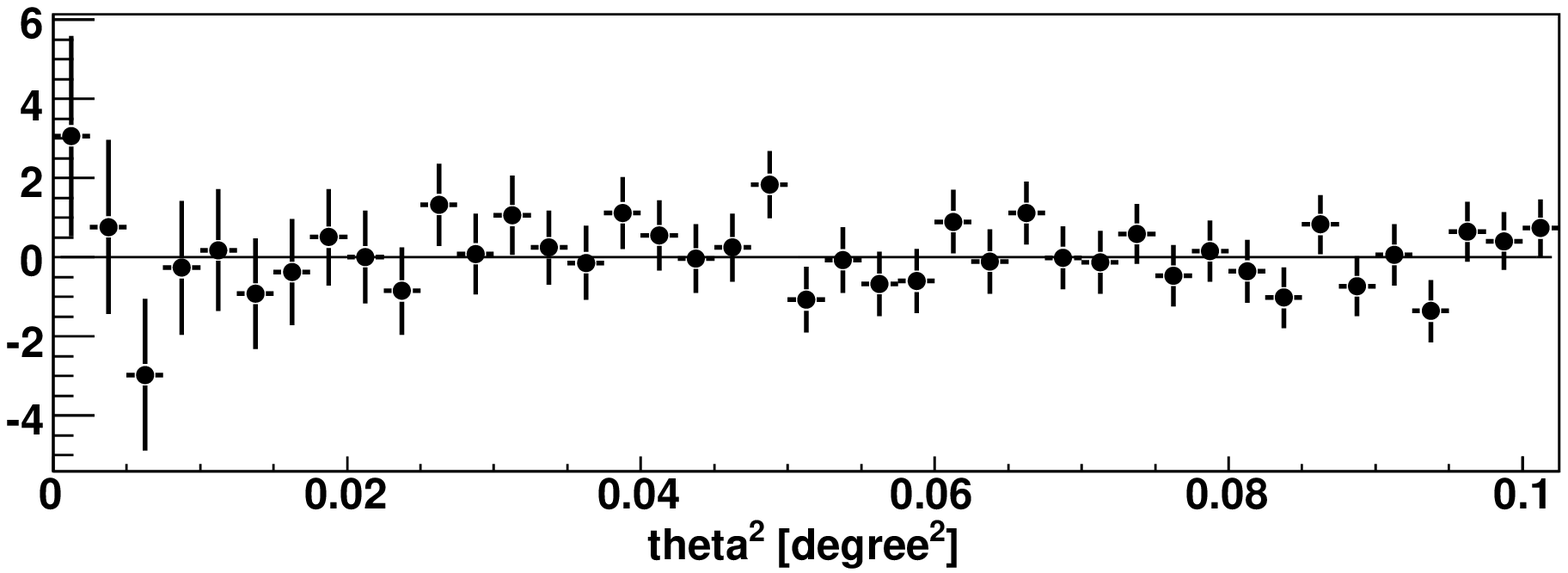}
  \caption{Comparison of the excess $\theta^2$ distribution for 38h of data from Mrk~421 (red circles) and a point-like source (blue triangles, 43h of Crab Nebula data) (upper panel) and the difference of both distributions (lower panel).}
  \label{fig_mrk421}
\end{figure}

$\theta^2$ distributions obtained for Mrk 501 and Mrk 421 are presented in fig.~\ref{fig_mrk501} and fig.~\ref{fig_mrk421}.
% bw
%% \begin{figure}%%%%%%%%%%%%%%%%%%%%%%%%%%%%%%%%%%
%%   \centering
%%   \includegraphics[width=8cm]{mrk501_comp.eps}
%%   \includegraphics[width=8cm]{mrk501_diff.eps}
%%   \caption{Comparison of the excess $\theta^2$ distribution for 26h of data from Mrk~501 (circles) and a point-like source (triangles, 17h of Crab Nebula data) (upper panel) and the difference of both distributions (lower panel).}
%%   \label{fig_mrk501}
%% \end{figure}
%% \begin{figure}
%%   \centering
%%   \includegraphics[width=8cm]{mrk421_comp_h.eps}
%%   \includegraphics[width=8cm]{mrk421_diff_h.eps}
%%   \caption{Comparison of the excess $\theta^2$ distribution for 38h of data from Mrk~421 (circles) and a point-like source (triangles, 43h of Crab Nebula data) (upper panel) and the difference of both distributions (lower panel).}
%%   \label{fig_mrk421}
%% \end{figure}%%%%%%%%%%%%%%%%%%%%%%%%%%%%%%%%%%
% color
%%%%%%%%%%%%%%%%%%%%%%%%%%%%%%%%%%

For both sources the $\theta^2$ distributions match with a corresponding point-like distribution.
Using the first 12 bins, which contain most of the excess events we calculated $\chi^2/n_{dof}$=7.0/11 (for Mrk~421) and 3.8/11 (for Mrk~501). 

\begin{figure*}%%%%%%%%%%%%%%%%%%%%%%%%%%%%%%%%%%
  \centering
  \includegraphics[width=0.49\textwidth]{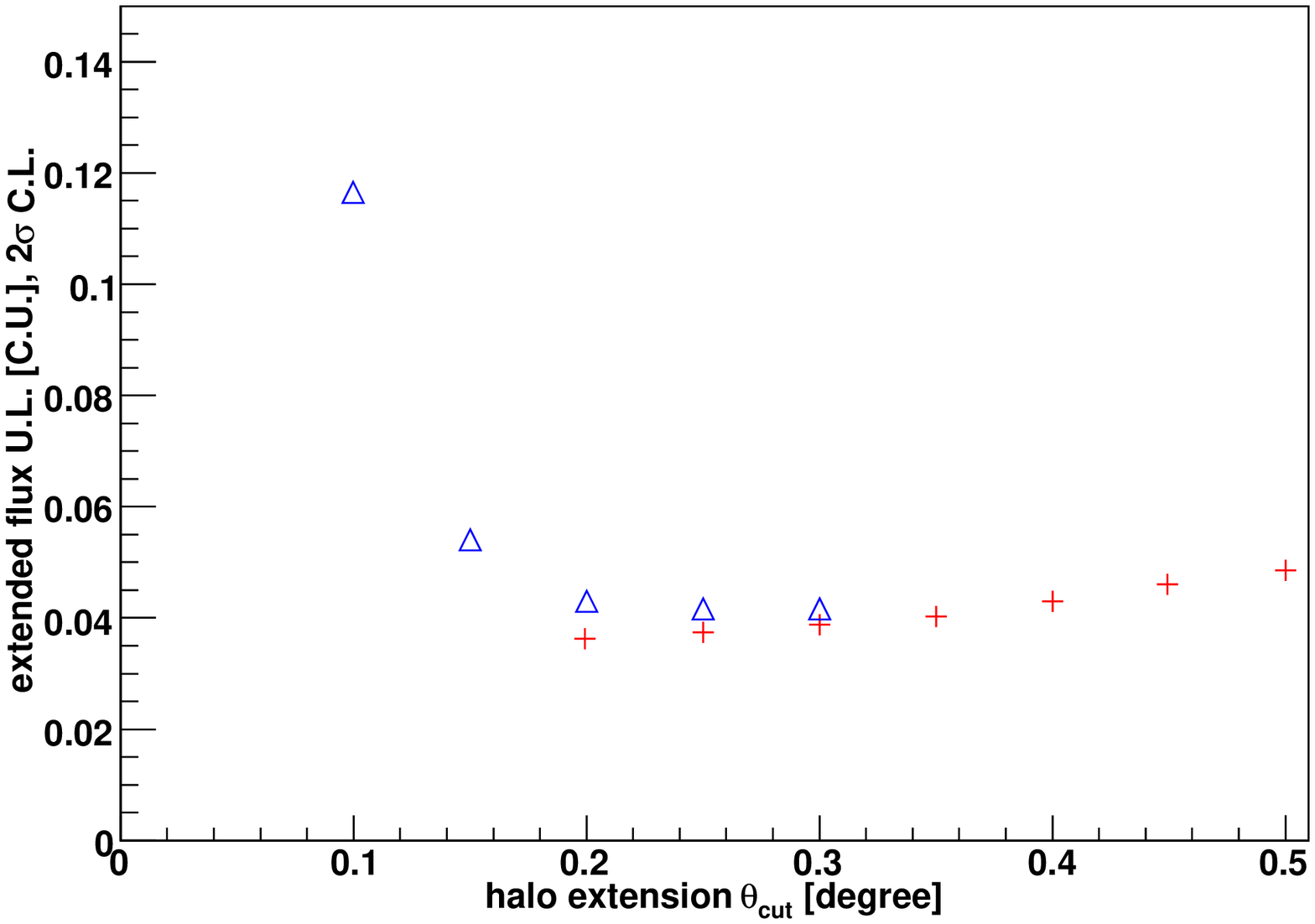}
  \includegraphics[width=0.49\textwidth]{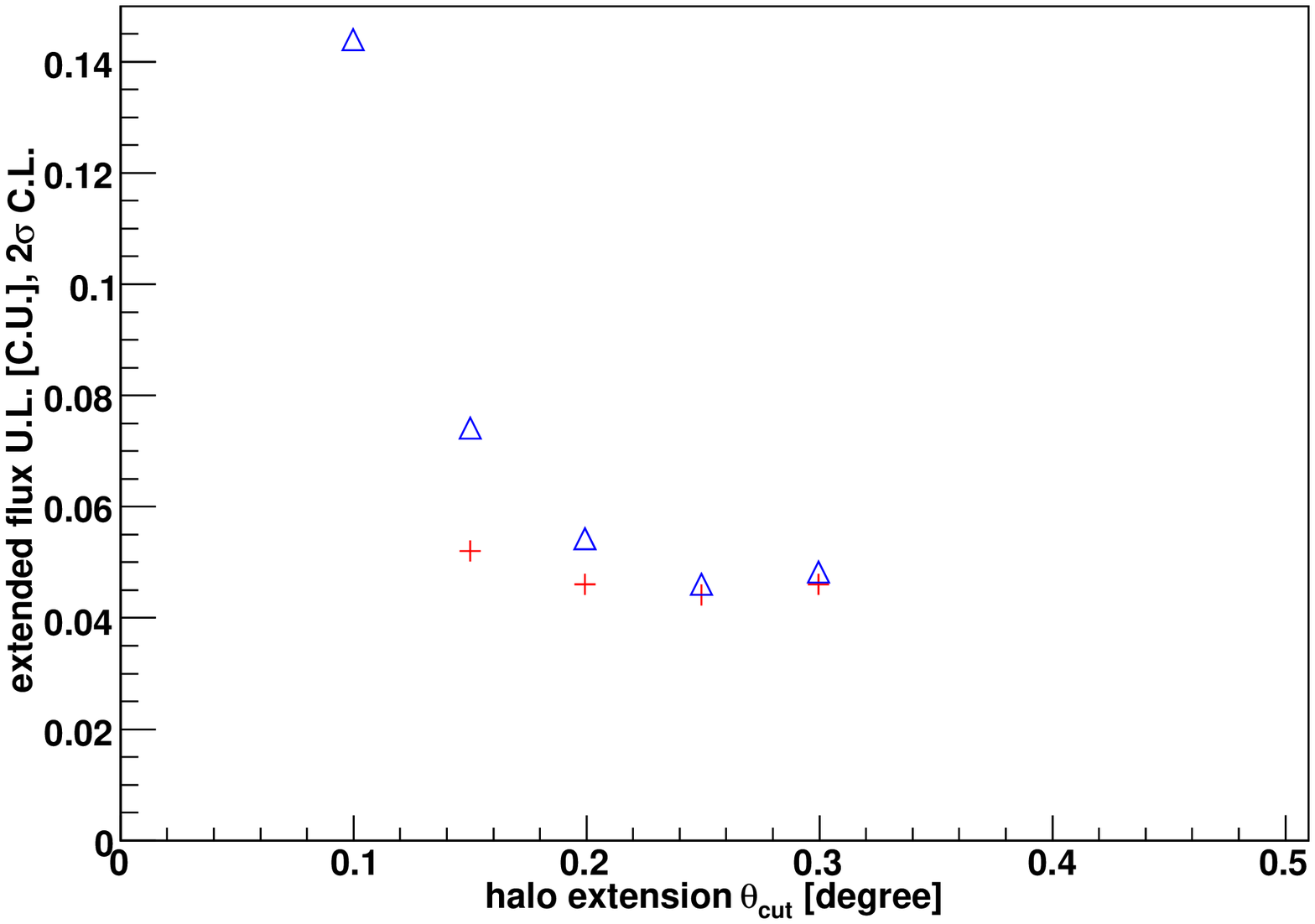}
  \caption{
Upper limit on the flux ($E>300$ GeV) of the extended emission from Mrk 501 (left figure) and Mrk 421 (right figure) in the C.U. 
for different source profiles and extensions $dN/d\theta \propto \theta^\beta$: 
%$0^\circ < \theta < \theta_{cut}$: flat disk ($\beta=1$) (open triangles), $0.1^\circ < \theta < \theta_{cut}$: $\beta=0$ (stars), $\beta=-1$ (crosses), $\beta=-2$ (dots).
$0^\circ < \theta < \theta_{cut}$: flat disk ($\beta=1$) (blue triangles); $0.1^\circ < \theta < \theta_{cut}$: $\beta=-1$ (red crosses). See more details in the text.
}
  \label{fig_uplim}
\end{figure*}

The sensitivity for the detection of an extended emission depends on the extension size and profile.
In our calculations we assumed a power-law profile of the emission ($dN/d\theta \propto \theta^\beta$) with various steepness indices $\beta=1, 0, -1, -2$.
We performed the calculations for extended emission radii up to 0.5$^\circ$ for Mrk~501. 
In the case of Mrk~421, the data were taken in wobble mode which allowed the estimation of the background from the same data set.
We calculated the upper limits for the extension radii up to 0.3$^\circ$.
In this case there is no overlap between the signal and the background regions in the wobble mode observations.
In fig.~\ref{fig_uplim} we present upper limits on the flux of the extended emission calculated for different extension radii and profiles for Mrk~501 (left figure) and Mrk~421 (right). 
%An upper limit on the flux of the extended emission calculated for different extension radii and profiles is shown in fig~\ref{fig_uplim}. 
% bw
%% \begin{figure*}%%%%%%%%%%%%%%%%%%%%%%%%%%%%%%%%%%
%%   \centering
%%   \includegraphics[width=8cm]{mrk501_uplim.eps}
%%   \includegraphics[width=8cm]{mrk421_uplim.eps}
%%   \caption{
%% Upper limit on the flux ($E>300$ GeV) of the extended emission from Mrk 501 (left figure) and Mrk 421 (right figure) in the C.U. 
%% for different source profiles and extensions $dN/d\theta \propto \theta^\beta$: 
%% %$0^\circ < \theta < \theta_{cut}$: flat disk ($\beta=1$) (open triangles), $0.1^\circ < \theta < \theta_{cut}$: $\beta=0$ (stars), $\beta=-1$ (crosses), $\beta=-2$ (dots).
%% $0^\circ < \theta < \theta_{cut}$: flat disk ($\beta=1$) (open triangles); $0.1^\circ < \theta < \theta_{cut}$: $\beta=-1$ (crosses). See more details in the text.
%% }
%%   \label{fig_uplim}
%% \end{figure*}%%%%%%%%%%%%%%%%%%%%%%%%%%%%%%%%%%
% color
%%%%%%%%%%%%%%%%%%%%%%%%%%%%%%%%%%

\begin{figure}%%%%%%%%%%%%%%%%%%%%%%%%%%%%%%%%%%
  \centering
  \includegraphics[width=8cm]{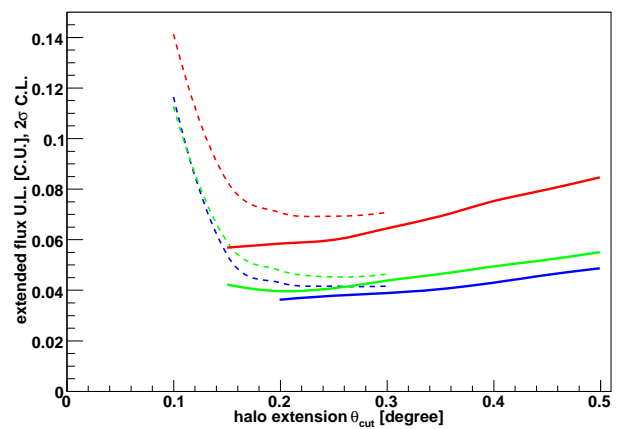} 
  \caption{
Upper limit (in C.U.) on the flux of the extended emission from Mrk~501 for different energy thresholds, 
different source profiles and extensions 
$dN/d\theta \propto \theta^\beta$:    
$0^\circ < \theta < \theta_{cut}$: flat ($\beta=1$) (dashed lines), $0.1^\circ < \theta < \theta_{cut}$, $\beta=-1$ (solid lines).
$E_{th}=1000$ GeV (red), 600 GeV (green) and 300 GeV (blue).}
  \label{fig_uplim_en}
\end{figure}

Flux upper limits for an extended emission from Mrk~501 for different energy thresholds are shown in fig.~\ref{fig_uplim_en}.
Depending on the source profile the upper limit for the extension radius $\sim 0.2^\circ$ ranges from $\sim$ 4\%C.U. above 300 GeV to 6-7\%C.U. above 1 TeV. 

% bw
%% \begin{figure}%%%%%%%%%%%%%%%%%%%%%%%%%%%%%%%%%%
%%   \centering
%%   \includegraphics[width=8cm]{ul_energies_bw.eps} 
%%   \caption{
%% Upper limit (in C.U.) on the flux of the extended emission from Mrk~501 for different energy thresholds, 
%% different source profiles and extensions 
%% $dN/d\theta \propto \theta^\beta$:    
%% $0^\circ < \theta < \theta_{cut}$: flat ($\beta=1$) (thin lines), $0.1^\circ < \theta < \theta_{cut}$, $\beta=-1$ (thick lines).
%% $E_{th}=1000$ GeV (solid lines), 600 GeV (dashed) and 300 GeV (dotted).}
%%   \label{fig_uplim_en}
%% \end{figure}%%%%%%%%%%%%%%%%%%%%%%%%%%%%%%%%%%
% color
%%%%%%%%%%%%%%%%%%%%%%%%%%%%%%%%%%

\section{Discussion}
As can be seen in fig.~\ref{fig_uplim} the most stringent upper limits for the halo search for $E_\gamma\simeq 300$~GeV are achieved  for a source extension of 0.2$^\circ$ -- 0.25$^\circ$ for Mrk~501 (4\% C.U.) and 0.2$^\circ$ -- 0.3$^\circ$ for Mrk~421 (5\% C.U.).
We checked that the values of upper limits for the emission profile $dN/d\theta \propto \theta^\beta$, $0.1^\circ < \theta < \theta_{cut}$, $\beta=-2$ or $\beta=0$ are nearly the same as for the case of $\beta=-1$.

For both sources, the best upper limits on the extended source flux are at a level of about 30\% of the quiescent point source flux.
Also, it is interesting to note that for an extension size $> 0.2^\circ$ one can observe only a marginal dependence on the emission profile.
Extension sizes $< 0.2^\circ$, becoming comparable to the telescope's PSF, provide worse upper limits. 

The non-detection of extended emission around Mrk 501 and Mrk 421 in the 0.3--1~TeV energy range imposes restrictions on the properties of the highest-energy $\gamma$-ray emission from these sources and/or on the physical characteristics of the intergalactic medium around the sources. 
Cascade photons with an energy of $E_\gamma\simeq 300$~GeV result from absorption of the primary $\gamma$-rays in the energy range $E_{\gamma_0}\simeq  20\left[E_\gamma/0.3\mbox{ TeV}\right]^{1/2}$~TeV which propagate over the distance $D_\gamma\simeq 40\kappa\left[E_{\gamma_0}/20\mbox{ TeV}\right]$~Mpc \citep{neronov09}.
Here $\kappa \sim 1$ is a numerical factor which accounts for uncertainty of the  EBL models, which give $0.6< \kappa <2.5$ for \cite{Kneiske:2003tx, Stecker:2005qs, Primack:2008nw, Franceschini:2008tp}.

Assuming that the EGMF strength is  much higher than $B\sim 10^{-12}$~G, as in the halo model of \citet{aharonian_halo}, the derived constraint on the extended source fluxes could be used to constrain the isotropic primary source power at energies above $\simeq 20$~TeV. Such a constraint is especially interesting in the view of the recent discovery of TeV $\gamma$-ray emission from the nearby radio galaxies \citep{m87_hess,m87_magic,m87,cena}. Within the general AGN unification scheme \citep{urry},  the high energy peaked BL Lacs (HBL), like Mrk 421 and Mrk 501, are believed to be the relativistically beamed versions of FR I type radio galaxies, like M87 and Cen A. 
Following the logic of the AGN unification scheme, the detection of TeV $\gamma$-ray emission from M87 and Cen A indicates that HBLs produce both beamed and isotropic TeV emission. 
Of course, the existence of such an isotropic component of VHE $\gamma$-ray  emission from Mrk 421 and/or Mrk 501 is difficult to verify, because the isotropic emission would produce a much smaller contribution to the 0.1-1~TeV band point source emission. 
However, if the isotropic emission spectrum of Mrk 421 and/or Mrk 501 extends, similarly to M87 \citep{m87_hess}, to energies $E_{\gamma_0}\ge 20$~TeV, absorption of the isotropically emitted $\gamma$-rays on the EBL leads to the production of an extended emission halo around HBL. 
The detection of an extended halo as discussed by \cite{aharonian_halo} would, therefore, provide a direct evidence for the existence of an isotropic multi-TeV emission from HBLs. 

In order to derive constraints on the isotropic emission from Mrk 501 and Mrk 421 from the limits on the extended emission flux, one needs to estimate the fraction of the halo flux within the measurement region of the radius $0.1^\circ-0.5^\circ$.  
The observable angular size of the halos around Mrk 421 and Mrk 501 (both at the redshifts $z\simeq 0.03$ and distance $D\sim 150$~Mpc) at the energy $E_\gamma\simeq 300$~GeV is expected to be $\Theta\sim D_\gamma/D\simeq 15^\circ\kappa$. 
Assuming a surface brightness profile $dN_\gamma/d\Omega\sim 1/\theta$ like in \cite{aharonian_halo}, one could find that the region $\theta\le 0.5^\circ$ around the source contains $\sim 3$\% of the halo emission. The halo is expected to be more compact at 1~TeV, $\Theta\simeq 8.3^\circ\kappa$, so that the central $\theta<0.5^\circ$ region contains  $\sim 6\%$ of the halo flux. This means that the total isotropic luminosity of Mrk 421 and Mrk 501 is limited to be less than 
$L_{\rm halo}(E_{\gamma_0}>20\mbox{ TeV})\simeq 1.3\kappa L_{\rm Crab}(E>0.3\mbox{ TeV})$ 
and $L_{\rm halo}(E_{\gamma_0}>36\mbox{ TeV})\simeq 1.3\kappa L_{\rm Crab}(E>1\mbox{ TeV})$.
It is clear that this limit is not very restrictive. 
Our analysis shows that the isotropic luminosity of HBLs could be more efficiently constrained via the search for an extended emission from a more distant source (for which the angular size of extended halos is smaller) and/or via observations with wider field of view instruments.

If the EGMF is much weaker than $10^{-12}$~G as already mentioned, deflections of cascade $e^+e^-$ pairs are not strong enough to isotropize cascade $\gamma$-ray emission. In this case no isotropically emitting halo around the point source is formed.  
The extended cascade source should appear more compact depending on the EGMF strength. 
Non-detection of extended emission constrains possible range of EGMF strength.  
In this short discussion we mention only limits on the EGMF with large correlation length. 
These limits could be extended to the case of arbitrary correlation length of the EGMF in a straightforward way, using the formalism of \citet{neronov09}.

The non-detection of extended emission at $E_\gamma\sim 300$~GeV could impose a bound on the EGMF only in the case when the time-averaged primary (beamed) source emission spectrum extends to the energies above $E_{\gamma_0}\ge 20$~TeV (see above). Unfortunately, the specific of observations in the TeV band and extreme variability of TeV-emitting blazars do not allow one to derive the time averaged spectra of the sources.

In the case of Mrk 421, a high energy cut-off in the spectrum at $E_{\rm cut}\simeq 1-5$~TeV has been repeatedly reported \citep{whipple_mrk421} (see however, \citet{konopelko08}). If such a cut-off is intrinsic and present in the time-averaged spectrum, the source luminosity at 20~TeV is expected to be  a factor of $\ge 15$ lower than the luminosity at 300~GeV (assuming that the intrinsic power law photon index is $\Gamma\simeq -1.7$, close to the one measured by {\it Fermi} \citep{fermi}). In this case the flux of the extended cascade emission at the energies $\ge 300$~GeV is a factor $\ge 15$ lower than the point source flux. This is consistent with the upper limits derived above. 
In the scenario of intrinsic cut-off in the Mrk~421 energy spectrum the upper bound on extended emission around Mrk 421, derived from MAGIC observations, does not constrain the strength of the EGMF.

On the contrary, the time-averaged spectrum of Mrk 501 extends, most probably to much higher-energies. 
No intrinsic high energy cut-offs in the low or high activity state of the source were reported. 
This implies that the source luminosity at the energies above 20 TeV can be of the same order as the luminosity at 300 GeV. 
At the same time, the bound on the extended source flux at a level of $\sim 0.04$ C.U. at 300~GeV is by a factor $\simeq 4$ lower than the point source flux at the same energy. 
If the intrinsic source flux at $\sim$20~TeV energy will prove to be higher than one forth of the flux at 300 GeV, the bound on the extended source flux, derived from MAGIC observations, might impose constraints on the strength of the EGMF within the region $D\sim D_\gamma\sim 40\kappa$~Mpc around Mrk 501. 
In fact, the spectrum of Mrk~501 from the analyzed data sample after correction for the absorption by using the \citet{Franceschini:2008tp} model (which provides a relatively low level of absorption) has a rather hard spectral index of $-2.24$.
This yields only a factor $(20~\mathrm{TeV}/0.3~\mathrm{TeV})^{0.24}=2.7$ decrease in the SED from the energy of 0.3 to 20~TeV. 
However both the uncertainties of the EBL absorption models and the mere fact that Mrk~501 is known to be variable tells us that the time-averaged spectrum over a longer periods of time may be found, for example, softer than what has been observed for the used data sample. 
Measurement of the intrinsic time-averaged source flux at 20 TeV will be possible only after precise measurements of the EBL in the mid-infrared and regular monitoring of the source on year(s) time scale. 
Therefore we give below only a qualitative estimate of the range of magnetic fields which might be constrained by the Mrk 501 data.

Assuming that the correlation length of the EGMF is much larger than the inverse Compton energy loss distance of electrons with energies $E_e\sim E_{\gamma_0}/2\sim 10$~TeV,  $D_e\simeq 30\left[E_e/10\mbox{ TeV}\right]^{-1}$~kpc, one can find that the size of the extended source around Mrk 501 is expected to be $\Theta_{\rm ext}\simeq 0.4^\circ\left[\tau/3.5\right]^{-1}\left[E_\gamma/300\mbox{ GeV}\right]^{-1}\left[B/10^{-14}\mbox{ G}\right]$ where $\tau=D/D_\gamma$ is the optical depth for the primary $\gamma$-rays with respect to the pair production on the EBL \citep{neronov09}. 
The non-detection of an extended source with $0.15^\circ<\Theta_{\rm ext}<0.5^\circ$ at an energy  $300$~GeV might constrain EGMF with a strength in the range $4\times 10^{-15}<B<1.3\times 10^{-14}$~G. 
A significant secondary cascade emission at 1~TeV will be possible if there is no cut-off in the primary spectrum below 40~TeV.
In this case non-detection of the extened emission at 1~TeV can be used for excluding EGMF strengths up to $B\simeq 8\times 10^{-14}$~G

It is very desirable to increase the sensitivity of the halo search.
An efficient way to enhance the sensitivity will be to use much longer observations, albeit if the systematics (e.g. the spot size, pointing accuracy) is well under control. 
For example if one would use 400h of on-source time (during 1 year, acceptable zenith angles) then a simple scaling as a square root of time will provide a sensitivity on the level of ~1\% C.U. 

Better sensitivity for a halo search could be achieved when the observed source is in a low emission state. 
In the high emission state the extended component remains constant, while the point-like emission provides additional background, hence deteriorating the sensitivity. 
This effect is especially important for the sources that have an extension comparable to the angular resolution of the telescope. 

Improving the angular resolution of the instrument is an efficient way of enhance the sensitivity for the halo search, especially when the emission radius is small. 
It would help to disentangle the extended component from the primary, point-like component down to lower extensions. 
For example with the planned CTA-like telescope array one could achieve an angular resolution in the range of 2 arcmin \citep{Hillas1989}. 
Accordingly 3 times smaller halo extension compared to the current study could be probed. 

By measuring with an instrument one order of magnitude more sensitive than MAGIC (like CTA) it seems realistic to achieve a sensitivity of $\sim$ 0.1 \% Crab for the halo search. 
This is true for sources with an extension $> 0.1-0.2^\circ$. 
An observation time of a few hundred hours will be necessary for such a study.

\section{Conclusions}
We have developed a novel method based on the Random Forest algorithm for the estimation of the source position.
This method improves the angular resolution by $\sim 20-30\%$. 

In the used data samples consisting of 26h observations of Mrk~501 and 38h of Mrk~421 no extended emission has been detected around these sources.
Our study showed that if there is an extended emission around Mrk 501, then its flux is $<$ 4\% C.U. (see fig.~\ref{fig_uplim} left), for the analysis energy threshold of 300 GeV. 
For Mrk 421 the upper limits are less stringent, because the source was in a high emission state, which causes the tail of the point-like emission to extend into the halo region thus creating additional background. 
The constraint on the extended emission flux from Mrk~421 is $< 5\%$ C.U. (see fig.~\ref{fig_uplim} right).

We analyzed different types of systematic errors connected with the observations of extended sources. 
We found them to be negligible compared to the upper limits obtained on the flux of the extended emission component of Mrk 421 and Mrk 501. 

With the second telescope, the MAGIC system can be operated in the stereo mode. 
By combining simultaneous information from both telescopes, the angular resolution will be improved. 
This will allow us to perform a search for the AGN halo with a better sensitivity.
The stereo observations improve the sensitivity 2.5 times at lower energies ($\sim 100$  GeV). 
Moreover, the better signal-to-noise ratio reduces the possible bias due to systematic errors. 
It will allow us to perform a sensitive search for halos at lower energies, thus extending the range of the EGMF strengths investigated with this method. 

A future $\gamma$-ray telescope project - the Cherenkov Telescope Array is aiming for a much better angular resolution.
In this case lower halo extensions can be investigated with an improved sensitivity.

%\end{document}

\begin{acknowledgements}
We would like to thank the Instituto de Astrofisica de 
Canarias for the excellent working conditions at the 
Observatorio del Roque de los Muchachos in La Palma. 
The support of the German BMBF and MPG, the Italian INFN,
the Swiss National Fund SNF, and the Spanish MICINN is gratefully acknowledged. 
This work was also supported by the Polish MNiSzW Grant N N203 390834, 
by the YIP of the Helmholtz Gemeinschaft, and by grant DO02-353
of the the Bulgarian National Science Fund.
\end{acknowledgements}

\end{document}